\documentclass[prd,aps,nofootinbib,eqsecnum,notitlepage,nobibnotes,superscriptaddress,preprintnumbers,
floatfix,twocolumn]{revtex4-1}
\usepackage{subfig}
\usepackage[titletoc,toc,title]{appendix}
\usepackage{mathtools}
\usepackage{amsmath, amsfonts,bm,amssymb}
\usepackage{graphicx}
\usepackage{xcolor}
\definecolor{darkerred}{rgb}{0.9, 0.17, 0.31}
\usepackage[a4paper, left=20mm,right=20mm,top=20mm,bottom=20mm]{geometry}
\usepackage[colorlinks=true,hyperfootnotes=true, linkcolor=darkerred, citecolor=cyan, urlcolor=magenta]{hyperref}
\usepackage[normalem]{ulem}
\def\dd{\mathrm{d}}
\newcommand{\be}{\begin{equation}} 
\newcommand{\ee}{\end{equation}}

\begin{document}

\title{Wormholes with matter haunted by conformally coupled ghosts}

\author{Bruno J. Barros}
\email{bjbarros@fc.ul.pt}
\affiliation{Instituto de Astrof\'isica e Ci\^encias do Espa\c{c}o,\\ 
Faculdade de Ci\^encias da Universidade de Lisboa,  \\ Campo Grande, PT1749-016 
Lisboa, Portugal}
\affiliation{Cosmology and Gravity Group, Department of Mathematics and Applied Mathematics, University of Cape Town, Rondebosch 7700, Cape Town, South Africa}
\author{\'Alvaro de la Cruz-Dombriz}
\email{alvaro.dombriz@usal.es}
\affiliation{Cosmology and Gravity Group, Department of Mathematics and Applied Mathematics,
University of Cape Town, Rondebosch 7700, Cape Town, South Africa}
\affiliation{ Departamento de F\'isica Fundamental, Universidad de Salamanca,
    P. de la Merced, 37008 Salamanca, Spain}

\author{Francisco S.N. Lobo}
\email{fslobo@fc.ul.pt}
\affiliation{Instituto de Astrof\'isica e Ci\^encias do Espa\c{c}o,\\ 
Faculdade de Ci\^encias da Universidade de Lisboa,  \\ Campo Grande, PT1749-016 
Lisboa, Portugal}
\affiliation{Departamento de F\'{i}sica, Faculdade de Ci\^{e}ncias, Universidade de Lisboa, Edifício C8, Campo Grande, PT1749-016 Lisbon, Portugal}
    
\date{\today}

\begin{abstract}
In this work, we present novel analytical solutions for static and spherically symmetric wormhole geometries threaded by an anisotropic distribution of matter conformally coupled to a scalar ghost field. We explore the main features of the theory, such as the dynamics of the scalar field and matter throughout the wormhole, as well as the role played by the nonminimal coupling. Furthermore, coupled ghosts in the presence of a scalar potential are considered and traversability conditions are analyzed within such geometrical scheme. More specifically, we find analytical solutions in which, although the energy density of the ghost is strictly negative, the energy density of matter may attain positive values.
\end{abstract}

\maketitle

\section{Introduction}\label{sec:intro}

In wormhole physics, a crucial element lies in the flaring-out condition of the throat. This condition, together with the Einstein field equations in standard General Relativity (GR), leads to the violation of the null energy condition (NEC) \cite{Morris:1988cz,Morris:1988tu}. Matter that violates the NEC has been denoted as {\it exotic matter}, and a number of wormhole solutions supported by these exotic fields have been analyzed in the literature \cite{Visser:1995cc,Alcubierre:2017pqm}. In this realm,  phantom fields that are possible candidates for the present accelerated expansion of the Universe have been explored \cite{Lobo:2005us,Sushkov:2005kj,Martinez:2020hjm}.  This phantom  field possesses an equation of state of the form $\omega= p/\rho <-1$, where $p$ is an isotropic cosmological pressure and $\rho$ is the energy density, which consequently violates the NEC.
Thus, this cosmic fluid, being the essential component for sustaining traversable wormholes, offers a natural framework for the presence of these exotic geometries.
Despite that, in a cosmological context, these cosmic fluids \cite{Lobo:2005yv,Lobo:2005vc,Lobo:2006ue,Lobo:2012qq,Bouhmadi-Lopez:2014gza} are considered homogeneous, gravitational instabilities may induce inhomogeneities. Hence, it is conceivable that these solutions originate from density perturbations within the cosmological background. Within the inflationary model of the early Universe, it has been proposed that macroscopic wormholes could naturally emerge from the submicroscopic structures that initially permeated the quantum foam \cite{Roman:1992xj}.

An interesting avenue of research to explore wormhole geometries is the context of modified theories of gravity, where the gravitational field equations usually involve higher-order curvature terms. This is of particular interest, as in addition to obtaining a richer class of solutions than in GR, subtle issues arise when concerning the energy conditions \cite{Capozziello:2013vna,Capozziello:2014bqa}. In fact, the classical energy conditions arise when one invokes the Raychaudhuri equation for the expansion of volume elements, where the specific term, $R_{\mu\nu}k^\mu k^\nu$, with $R_{\mu\nu}$ the Ricci tensor and $k^\mu$ a null vector, is considered. The positive character of $R_{\mu\nu}k^\mu k^\nu$ entails the attractive nature of gravity. Thus, in GR, through the Einstein field equations, this implies $T_{\mu\nu}k^\mu k^\nu \geqslant 0$, which is precisely the definition of the NEC \cite{Hawking:1973uf,Visser:1995cc,Alcubierre:2017pqm}. In modified theories of gravity, in order to replace the Ricci tensor using the corresponding field equations, one arrives at more complicated conditions (see Refs. \cite{Capozziello:2013vna,Capozziello:2014bqa,Albareti:2012se,Albareti:2012va} for more details). Therefore, within this framework, it has been demonstrated explicitly that wormhole geometries can be theoretically formed without requiring exotic matter, at the expense of relying on a modified gravity approach \cite{Harko:2013yb}. In that reference, it was shown that the matter threading the wormhole satisfies all of the energy conditions, while the higher-order curvature terms, resembling a gravitational fluid, play the primary role of sustaining these unconventional wormhole geometries. Indeed, this is an active area of research and wormhole geometries have been explored in a plethora of modified theories of gravity, namely: $f(R)$ gravity \cite{Lobo:2009ip,DeBenedictis:2012qz,Pavlovic:2014gba,Bahamonde:2016ixz,Eiroa:2015hrt,Samanta:2019tjb,Godani:2019kgy,Rosa:2023olc,DeFalco:2021ksd,DeFalco:2021klh}; nonminimal curvature-matter couplings \cite{Garcia:2010xb,MontelongoGarcia:2010xd,Bertolami:2012fz}; 
modified teleparallel gravity \cite{Boehmer:2012uyw,Jamil:2012ti,Bahamonde:2016jqq,Sharif:2013exa}; the Palatini \cite{Lobo:2013adx,Lobo:2013vga,Lobo:2014fma,Lobo:2014zla,Olmo:2015dba,Olmo:2015bya,Bambi:2015zch,Bejarano:2016gyv,Menchon:2017qed,Lobo:2020vqh} and hybrid metric-Palatini approaches of gravity \cite{Capozziello:2012hr,Rosa:2018jwp,Rosa:2021yym}; $f(R,T)$ gravity \cite{Moraes:2017mir,Zubair:2016cde,Yousaf:2017hjh,Moraes:2017rrv,Moraes:2019pao,Mishra:2019jpa,Rosa:2022osy}; Einstein-Cartan gravity \cite{Mehdizadeh:2017tcf}; Eddington-inspired Born-Infeld gravity \cite{Harko:2013aya}; Einstein-Gauss-Bonnet gravity \cite{Mehdizadeh:2015jra}; conformal Weyl gravity \cite{Lobo:2008zu}; and Lovelock gravity \cite{KordZangeneh:2015dks}; among many other gravitational theories.

On another note, much work has also been explored within the framework of scalar-tensor theories (STTs) \cite{Bronnikov:1973fh,Bronnikov:2004ax,Bronnikov:2001ae,Bronnikov:2006qj,Bronnikov:1996de} (see also \cite{Barcelo:1999hq,Willenborg:2018zsv,Chew:2018vjp}). Within STTs it is known \cite{Bronnikov:2006pt,Bronnikov:2010tt} that there are no traversable wormhole solutions for a positive coupling (assuring the nonghost nature of the graviton) nor with a nonghost scalar field. Thus, in STTs of gravity, in order to reproduce wormhole solutions at least one ghost degree of freedom is required. Let us stress that although it is known that a ghost scalar field is unstable \cite{Nandi:2016ccg,Gonzalez:2008wd}, one may regard the field theory model at hand to be treated as an effective theory, valid up to a certain energy scale, which may be derived from a stable fundamental theory \cite{Nojiri:2003rz}.
Additionally, it has been shown that if one defines the effective cutoff scale at 100 MeV, the timescale of the instability can be greater than the age of the Universe \cite{Carroll:2003st}. Therefore, ghost scalar fields can still provide insights into wormhole physics. Lastly, ghost fields are still extensively studied to this day due to their historical context: they were one of the first proposals in 1973, as an exotic fluid to sustain a wormhole, or, as originally dubbed in Ref.~\cite{Ellis:1973yv}, a {\it{drainhole}}. This is one of the reasons that ghost scalar fields, {\it i.e.}, a scalar field with a negative kinetic term, are widely used in the literature as the source of supporting wormhole, and other nontrivial, geometries \cite{Karakasis:2021tqx,Bronnikov:2018vbs,Bronnikov:2010hu,Lazov:2017tjs}. Accordingly, in this work we adopt the Ellis ghost scalar field, presented as the {\it plumber's best friend} in the seminal papers in Refs.~\cite{Ellis:1973yv} (see also \cite{Bronnikov:1973fh}). On the other hand, the ghost field is conformally coupled to matter and may also acquire a nonzero self-interacting potential. Consequently, in the following we shall explore the main dynamics and properties of the ghost field in such a setting while focusing on its interplay with the additional matter source. It is to be noted that the same procedure employed in this work to couple the scalar field to matter, through a conformal transformation, was already employed by Bronnikov in Ref.~\cite{Bronnikov:1973fh}. However, this was done by considering that the conformal transformation affects Maxwell's electromagnetic Lagrangian, which is conformally invariant, and thus no effective interaction arises.

Nonetheless, herein we look for solutions where the additional coupled matter source is nonvanishing and meticulously study the synergy between the scalar source and the anisotropic matter species. This work is organized in the following manner. In  Sec. \ref{sec:model}, we present the action and field equations for a ghost scalar field and a distribution of matter, in the Einstein frame, and briefly explore the violations of the energy conditions of the theory. In Sec. \ref{sec_massless}, we investigate solutions with a vanishing self-interacting potential and find specific analytical solutions. In Sec. \ref{sec_mass}, we analyze solutions with a nonvanishing scalar potential and briefly study the traversibility conditions of specific solutions obtained. Finally, in Sec. \ref{sec:conclusions}, we discuss our results and conclude.

\section{(Re)visiting ghosts: the model}\label{sec:model}

\subsection{Setting the stage}

This present work regards interactions betwen a canonical scalar field ${\phi:\mathcal{M}\rightarrow \mathbb{R}}$ and an anisotropic distribution of matter, threading a static and spherically symmetric wormhole geometry. One common approach to introduce a nonminimal coupling to matter is by assuming that matter experiences a different metric, say $\bar{g}$, from the one on which $\phi$ propagates, $g$. These two different metrics can be related through a Weyl scaling stemming from a conformal transformation. More specifically, given a Riemannian space $(\mathcal{M},g)$, consisting of a smooth manifold $\mathcal{M}$ endowed with a metric $g$, a conformal transformation $\zeta$ is a diffeomorphism
\begin{eqnarray}
\mathcal{M} &\xrightarrow{\quad\zeta\quad}& \mathcal{M} \nonumber\\
x^{\mu} &\xmapsto{\quad\,\,\,\,\quad} & \tilde{x}^{\mu}=\zeta(x^{\mu}) \nonumber
\end{eqnarray}
such that ${\zeta^*g=\Omega^2(x^{\mu})\,g}$, where $\zeta^*g$ denotes the pullback of $g$ by $\zeta$, {\it i.e.}, ${\zeta^*g = g\circ \zeta}$. It is a change of coordinates that leaves the metric invariant up to a conformal factor $\Omega^2$. Note that the trivial case $\Omega=1$ simply represents an isometry and we have assumed a squared function in order for the metric to preserve its signature. 

In STTs of gravity, the conformal factor is assumed to be mediated by a scalar field $\phi$ naturally present in the theory, {\it i.e.}, ${\Omega(x^{\mu})=\Omega[\phi(x^{\mu})]}$. It is then common to evoke a given theory ${\mathcal{S}\left[ g,\phi,\psi \right]}$ (where $\psi$, portraying the matter fields, is any set of sections of the frame bundle of $\mathcal{M}$) in one of two different representations or, as it is customary to say, two different frames, which portray the same underlying theory but with distinct interpretations for gravitational phenomena. One spacetime, say, ${(\mathcal{M},g)}$, is known as the Einstein frame if the scalar degree of freedom is minimally coupled to gravity, ${\mathcal{L}_g=\mathcal{L}_g(g)}$, but nonminimally coupled to matter, ${\mathcal{L}_m=\mathcal{L}_m(g,\phi,\psi)}$. In the other frame, ${(\mathcal{M},\tilde{g})}$, with ${\tilde{g}=\Omega^2 g}$, dubbed the Jordan frame, $\phi$ is minimally coupled to matter,  $\mathcal{L}_m=\mathcal{L}_m(g,\psi)$, and nonminimally coupled to gravity, $\mathcal{L}_g=\mathcal{L}_g(g,\phi)$. This nonminimal coupling to gravity, in the so-called Jordan frame, can be interpreted as endowed with a varying gravitational constant and was introduced by Brans and Dicke as an effort to integrate Mach's principle into the framework of GR \cite{Brans:1961sx,Dicke:1961gz,Brans:1962zz}. Note that both ${(\mathcal{M},g)}$ and ${(\mathcal{M},\tilde{g})}$ possess identical causal structure. We refer the reader to \cite{Fujii:2003pa,Faraoni:2004pi} to explore the details of scalar-tensor cosmology. One may then write the action for a given theory in a given frame and map it to the other simply by means of a Weyl scaling, ${g\mapsto \tilde{g}=\Omega^2 g}$. Nevertheless, there remains ongoing discussion regarding which frame should be regarded as the physical one \cite{Faraoni:2004pi}. 

\subsection{Action and field equations}

The total action for a ghost scalar field $\phi$ and a distribution of matter, collectively denoted by $\psi$, in the Einstein frame, is given by
\be\label{action}
\mathcal{S}= \int \omega_g \left(\frac{R}{2\kappa^2}+\frac{1}{2}g^{\mu\nu}\partial_{\mu}\phi\,\partial_{\nu}\phi -V\right) + \mathcal{S}_m\left(\tilde{g}_{\mu\nu},\psi\right)\,,
\ee
where ${\kappa^2=8\pi G}$; ${\omega_g=\dd^4x\sqrt{-g}}$ is the volume form; $g$ being the determinant of $g_{\mu\nu}$; $R$ is the standard curvature scalar constructed from the metric $g$; and $V=V(\phi)$ is the scalar potential. The positive sign in the kinetic term of the scalar field above characterizes its ghost nature \cite{Bronnikov:2004ax}. Note that the matter fields propagate in geodesics related to a metric $\tilde{g}$, conformally associated to $g$ through a Weyl scaling \cite{Fujii:2003pa,Faraoni:2004pi}, given above:
\be\label{weyl}
\tilde{g}_{\mu\nu}=\Omega^2(\phi)\,g_{\mu\nu}\,.
\ee 
In essence, the propagation of the matter fields is affected by the dynamics of $\phi$, as $\mathcal{S}_m$ depends on $\phi$ through Eq.~\eqref{weyl}.

Varying the action \eqref{action} with respect to $g^{\mu\nu}$ yields the following field equations:
\be
R_{\mu\nu}-\frac{1}{2}g_{\mu\nu}R = \kappa^2\,T^{( {\rm eff} )}_{\mu\nu}\,.
\label{EoM}
\ee 
$T^{( {\rm eff} )}_{\mu\nu}$ is the effective energy-momentum tensor defined in terms of the two matter sources within the theory,
\be\label{T_eff}
T^{( {\rm eff} )}_{\mu\nu}\, \coloneqq\, T^{( \phi )}_{\mu\nu} + T^{( m )}_{\mu\nu}\,,
\ee
where
\begin{eqnarray}
T^{(\phi)}_{\mu\nu} &=& -\partial_{\mu}\phi\,\partial_{\nu}\phi + g_{\mu\nu}\mathcal{L}_{\phi}
\label{Tphi}
\,, \\
T^{(m)}_{\mu\nu} &=& -\frac{2}{\sqrt{-g}}\frac{\delta\mathcal{S}_m }{\delta g^{\mu\nu}}
\label{Tm}
\end{eqnarray}
are the energy-momentum tensors of the ghost field and matter, respectively, with the ghost Lagrangian density,
\be\label{L_phantom}
\mathcal{L}_{\phi}=\frac{1}{2}\partial^{\mu}\phi\,\partial_{\mu}\phi -V\,.
\ee

The contracted Bianchi identities yield\sout{s} the following conservation relations for the energy-momentum tensors:
\begin{eqnarray}
\nabla_{\mu}T^{(\phi)}{}^{\mu}{}_{\nu} &=& -T^{(m)}\,\nabla_{\nu}\left( \ln \Omega\right)\,, \\
\nabla_{\mu}T^{(m)}{}^{\mu}{}_{\nu} &=& T^{(m)}\,\nabla_{\nu}\left( \ln \Omega\right)\,,
\end{eqnarray}
where $T^{(m)}$ is the trace of the matter energy-momentum tensor \eqref{Tm}. Since the conformal factor $\Omega$ is mediated by $\phi$ as per Eq.~\eqref{weyl}, we may write
\begin{eqnarray}
\nabla_{\mu}T^{(\phi)}{}^{\mu}{}_{\nu} &=& -\frac{\Omega_{\phi}}{\Omega}\, T^{(m)}\nabla_{\nu}\phi\,, 
\label{SETfield}
\\
\nabla_{\mu}T^{(m)}{}^{\mu}{}_{\nu} &=& \frac{\Omega_{\phi}}{\Omega}\, T^{(m)}\nabla_{\nu}\phi\,,
\label{SETm}
\end{eqnarray}
where ${\Omega_{\phi}=\dd\Omega/\dd\phi}$. Note that, while the effective  energy-momentum tensor \eqref{T_eff} is conserved, each individual species is not: {\it i.e.}, there is a flow of energy-momentum between matter and the ghost mediated by the conformal factor ${\Omega(\phi)}$.  This is germane to the fact that we have considered the action \eqref{action} in the Einstein frame, where the scalar degree of freedom is decoupled from gravity but {\it nonminimally coupled to matter}. For obvious reasons, this specific form of interactions bears the nomenclature of {\it conformal couplings} \cite{Fujii:2003pa,Faraoni:2004pi}.

In this current work, we focus on the exponential case given by
\be
\Omega(\phi)=\text{e}^{-\kappa\beta\phi}\,,
\ee
where $\beta$ is a dimensionless constant governing the strength of the interaction between the two components. Thus, from Eqs.  (\ref{SETfield}) and (\ref{SETm}), we have
\be\label{interaction}
\nabla_{\mu}T^{(m)}{}^{\mu}{}_{\nu}=-\nabla_{\mu}T^{(\phi)}{}^{\mu}{}_{\nu}=-\kappa\beta\, T^{(m)}\nabla_{\nu}\phi\,.
\ee
This specific form of the coupling has been widely studied throughout the literature, for example in the context of dark energy driven by a scalar field conformally coupled to a dark matter component \cite{Amendola:1999er,Barros:2018efl}.
Thus, in the following we shall study the existence of wormhole geometries and their dynamical behavior as different values of the conformal coupling, $\beta$, are considered.

Finally, the variation of the action \eqref{action} with respect to $\phi$ provides the coupled Klein-Gordon equation,
\be\label{kleinG}
\square \phi + V_{\phi} = -\kappa\beta\, T^{(m)} \,,
\ee
where ${V_{\phi}=\dd V/\dd\phi}$ and ${\square=g^{\mu\nu}\nabla_{\mu}\nabla_{\nu}}$ is the d'Alembert operator.

\subsection{Wormhole metric}\label{wormhole:metric}

Let us examine a static and spherically symmetric wormhole geometry characterized by the following line element \citep{Morris:1988cz,DeFalco:2021btn,DeFalco:2020afv}:
\be\label{metric}
\dd s^2 = -\text{e}^{2\Phi(r)} \dd t^2 + \frac{\dd r^2}{1-\frac{b(r)}{r}} + r^2 \left( \dd \theta^2 + \sin^2 \theta \, \dd \varphi^2 \right)\,.
\ee
The radial coordinate $r$ possesses the range ${r \in [r_0,\infty[}$, where the minimum value $r_0$ is defined as the wormhole throat. The redshift function, $\Phi(r)$, encloses the effects of the gravitational redshift and tidal accelerations and is assumed to be finite in the entire domain of $r$ so that no event horizons are present, rendering the wormhole traversable.  The shape function, $b(r)$, portrays the geometrical shape of the wormhole and obeys the following conditions: $b(r_0)=r_0$; the flaring-out condition, {\it i.e.},  ${b'(r)<b(r)/r}$ (so that ${b'(r_0)<1}$ at the throat), near the throat, where the prime denotes a derivative with respect to the radial coordinate $r$; and ${b(r)<r}$. 
In principle, one can construct asymptotically flat spacetimes, in which ${b(r)/r \rightarrow 0}$ and ${\Phi(r) \rightarrow 0}$ as ${r \rightarrow +\infty}$.
Note that the $g_{rr}$ component of the metric entails a coordinate, albeit not a physical, singularity, at $r_0$, which can be avoided by a suitable change of coordinates. For instance, the proper radial distance ${l(r)= \pm \int_{r_0}^r \left[1-b(r)/r \right]^{-1/2}{\rm d}r}$ is required to be finite everywhere.

We consider that the ghost field couples to an anisotropic distribution of matter threading the wormhole and given by the following energy-momentum tensor:
\be
T^{(m)}_{\mu\nu} = (\rho_m+p_m)u_{\mu}u_{\nu}+p_mg_{\mu\nu} - (\tau_m+p_m)\chi_{\mu}\chi_{\nu},
\label{EM-matter}
\ee
where $u_{\mu}$ is the four-velocity vector; ${\chi^{\mu} = \delta^{\mu}_r \sqrt{1-b/r}}$ is the radial component of a unit spacelike vector; and the energy density, radial tension, and tangential pressure (orthogonal to the radial direction) are represented by $\rho_m$, $\tau_m$, and $p_m$, respectively. This can be summarized as
\be
T^{( m )}{}^{\mu}_{\;\nu}=\,\text{diag} (-\rho_m,\,-\tau_m,\,p_m,\,p_m)\,,
\ee
where the trace reads
\be\label{T_m_trace}
T^{(m)}=-\rho_m-\tau_m+2\,p_m\,.
\ee 
Note that the evolution of this distribution of matter throughout the wormhole spacetime depends explicitly on the ghost evolution, through the coupling relation \eqref{interaction}.

Given the symmetries of our spacetime continuum \eqref{metric}, we will assume the scalar source to be a function of the radial coordinate only, {\it i.e.}, ${\phi=\phi(r)}$. Within this geometrical setting, and taking into account Eq. (\ref{Tphi}), we identify the nonzero components of ${T^{( \phi )}_{\mu\nu}}$ as
\begin{eqnarray}
\rho_{\phi} &\,\coloneqq& -T^{( \phi )}{}^t{}_t = -\frac{1}{2}\left( 1- \frac{b}{r} \right){\phi'}^2 + V\,, \label{rho_phi}  \\
\tau_{\phi} &\,\coloneqq& -T^{( \phi )}{}^r{}_r  = \frac{1}{2}\left(1- \frac{b}{r} \right){\phi'}^2 +V\,, \label{tau_phi} \\
p_{\phi} &\,\coloneqq& T^{( \phi )}{}^{\theta}{}_{\theta} = T^{( \phi )}{}^{\varphi}{}_{\varphi} = T^{( \phi )}{}^t{}_t ,
\end{eqnarray}
respectively.
Without a loss of generality, henceforth, we set ${\kappa^2 = 1}$. The field equations \eqref{EoM} can now be written as
\begin{eqnarray}
\rho_{{\rm eff}} &=& \rho_{m}  + \rho_{\phi} = \frac{b'}{r^2}\,, \label{ee1} 	\\
\tau_{{\rm eff}} &=& \tau_{m}  + \tau_{\phi}= \frac{b}{r^3} - 2\left( 1-\frac{b}{r} \right)\frac{\Phi'}{r}\,, \label{ee2} \\
p_{{\rm eff}} &=& p_{m} + p_{\phi} = \left( 1-\frac{b}{r}\right)\left\{ \Phi'' +\left(\Phi'+\frac{1}{r}\right)\right.\times\nonumber \\
&\,&\hspace{1.7cm}\left.
\times
\left[\Phi'+\frac{b-b'r}{2r\left(r-b\right)}\right]\right\} . \label{ee3}
\end{eqnarray}
On the other hand, Eq. \eqref{kleinG} with the metric \eqref{metric} provides the following equation\sout{s} of motion for the $\phi$ field:
\begin{eqnarray}
\label{kleingordon}
\phi''\left( 1-\frac{b}{r} \right)&+&\frac{\phi'}{2r}\left[4-3\frac{b}{r}-b'+r\Phi'\left(1-\frac{b}{r}\right)\right]+V_{\phi} \nonumber \\
&=&-\beta \,T^{(m)}\,,
\end{eqnarray}
where we identify the scalar interaction to matter as the right-hand side of the above equation. Hence, the evolution of the ghost field will be intimately related with the dynamics of matter within the wormhole. Analogously, the conservation relation for matter, Eq.~\eqref{interaction}, provides the following continuity equation:
\be\label{continuity}
\tau_m'+\frac{2}{r}\left(\tau_m+p_m\right) +\Phi'\left(\tau_m-\rho_m\right)=\beta\phi'\,T^{(m)}\,.
\ee
Note that this equation is not independent, since it can be derived by combining the radial derivative of Eq.~\eqref{ee2} with Eqs. \eqref{ee1}, \eqref{ee3}, and \eqref{kleingordon}. In the absence of a coupling, by setting ${\beta=0}$, one recovers the standard energy-momentum \eqref{EM-matter} conservation.

Due to the nonlinearity of the field equations, it is extremely difficult to find solutions. Indeed, note that now one has four equations, namely, the field equations \eqref{ee1}--\eqref{kleingordon}, and seven unknown functions of $r$, {\it i.e.},   $\rho_m$, $\tau_m$, $p_m$, $\phi$, $V$, and the metric functions $b(r)$ and $\Phi(r)$. Thus, we are free to impose three assumptions, in order to close the system.

Taking into account the flaring-out condition, Eqs. (\ref{ee1}) and (\ref{ee2}) imply the following condition in the vicinity of the throat, $r=r_0$:
\be
\rho_{{\rm eff}} - \tau_{{\rm eff}} =  \rho_{m}  + \rho_{\phi} -  \tau_{m}  - \tau_{\phi} < 0\,.
\ee

\section{Solutions with a vanishing self-interacting potential}\label{sec_massless}

The case we consider throughout this section consists of a ghost field in the absence of a self-interacting potential, {\it i.e.}, ${V=0}$. In Secs.~\ref{uncoupled_phantom} and 
\ref{SectionIII.B} below, we consider a constant redshift function ${\Phi=\Phi_0}$ and the following shape function:
\be\label{shapefunc}
b=\frac{r_0^2}{r}\,.
\ee
Thus, resorting to the modified field equations \eqref{ee1}-\eqref{ee3}, the Klein-Gordon equation \eqref{kleingordon} can be written as
\be\label{kleingordon1}
\left( 1-\frac{r_0^2}{r^2} \right)\left[\phi''+\phi'\left(\frac{1}{r}-\beta\phi'\right)\right] +\frac{\phi'}{r}+2\beta\frac{r_0^2}{r^4}=0 \,.
\ee
%

\subsection{Noninteracting case: Ellis drainhole with matter}\label{uncoupled_phantom}

The uncoupled case ${\beta=0}$ significantly simplifies Eq.~\eqref{kleingordon1}, which yields
\be
\label{KG massless uncoupled}
{\phi}''\left( 1-\frac{r_0^2}{r^2} \right)+\frac{{\phi}'}{r}\left(2-\frac{r_0^2}{r^2}\right) =0 \,,
\ee
that provides the following solution for the ghost field, which is uncoupled to matter,
\be\label{uncoupled}
{\phi} = {\phi}_0+C\tan^{-1}\left(\sqrt{\frac{r^2}{r_0^2}-1}\,\right)\,,
\ee
where the parameter ${{\phi}_0= {{\phi}}(r_0)}$ is the value of the field at the throat, which together with $C$ consitutes the two integration constants issued from Eq. \eqref{KG massless uncoupled}. Far from the throat, we have 
\be\label{uncoupled_inf}
{\phi}_{\infty}=\lim_{r\rightarrow\infty}{{\phi}}={\phi}_0+\frac{C\pi}{2}\,.
\ee

Hence, we express the constant $C$ in a more natural way, as the difference of the values of the scalar field at the throat and at infinity, {\it i.e.},
\be\label{dif_uncoupled}
C=\frac{2}{\pi}\left({\phi}_{\infty}-{\phi}_0\right)\,,
\ee
so that Eq. \eqref{uncoupled} renders
\be
{\phi}={\phi}_0+\frac{2}{\pi}\left( {\phi}_\infty-{\phi}_0 \right)\tan^{-1}\left(\sqrt{\frac{r^2}{r_0^2}-1}\,\right)\,.
\ee
Taking the derivative of the above equation and plugging it into Eq.~\eqref{rho_phi} gives an energy density for the scalar field, given by
\be\label{rho_phi_uncoupled}
{\rho}_{\phi}=-\frac{r_0^2C^2}{2r^4}= -\frac{2r_0^2}{\pi^2r^4}\left({\phi}_0-{\phi}_\infty\right)^2\,.
\ee

Note that there is a subtle difference between the solutions presented in the literature, including the Ellis ghost \cite{Ellis:1973yv,Karakasis:2021tqx}, and this one, namely that we have freedom in the value of the integration constant $C$, which is directly associated with the ghost energy density as per \eqref{rho_phi_uncoupled} above and gives a whole new family of solutions (one for each value of $C$). This arises due to the fact that we are considering an extra matter source, whose energy density, ${\rho}_m$, together with ${\rho}_{\phi}$ reproduces ${{\rho}_{\rm eff}=b'/r^2}$ as given by Eq.~\eqref{ee1}.

It is straightforward to notice that, although the NEC for the ghost,
\be\label{nec1}
\rho_{\phi}-\tau_{\phi}=-\frac{C^2r_0^2}{r^4}\,,
\ee
is always violated, the same is not always true for the matter source,
\be
\rho_m-\tau_m = \frac{r_0^2}{2r^4}\left(C^2-2\right)\,,
\ee
where it is positive for $C^2>2$. Thus, by adding an extra ingredient to the theory, we are able to tune the difference of values of the ghost at the throat and at infinity ({\it i.e.}, the parameter $C$), to make the ordinary matter source nonexotic.
We follow to fix the value of the field at infinity to zero, ${\phi}_\infty=0$, rendering the ghost energy density as per Eq.~\eqref{rho_phi_uncoupled} to depend solely on the value of the field at the throat, ${\phi}_0$.

Let us notice two limiting cases at this stage: should no matter source be assumed besides the ghost, the value  $C=\sqrt{2}$ would be fixed by the first Einstein equation, Eq.~\eqref{ee1}. In this case the ghost is the only source supporting the wormhole, {\it i.e.}, ${\rho_{\rm eff} = \rho_{\phi}}$, and therefore we get $C^2=2$. This was already studied in the literature in Refs.~\cite{Ellis:1973yv,Karakasis:2021tqx}. Finally, in the trivial case when ${C=0}$, no ghost scalar field is present, and thus only matter threads the wormhole.
\begin{figure}[t]
    \centering
    \includegraphics[scale=0.415]{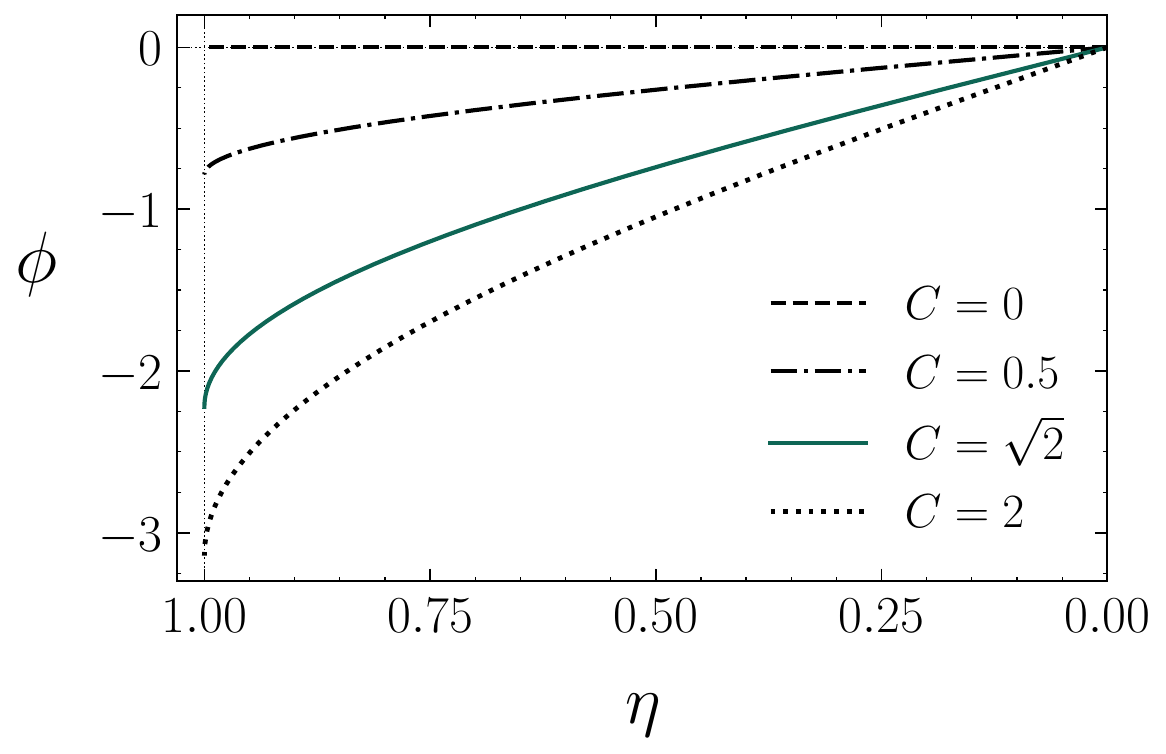}
    \caption{Ghost field profile, Eq.~\eqref{uncoupled}, for the uncoupled case with different values of $C$.}
    \label{phi_un}
\end{figure}

\begin{figure}[t]
    \centering
    \includegraphics[scale=0.45]{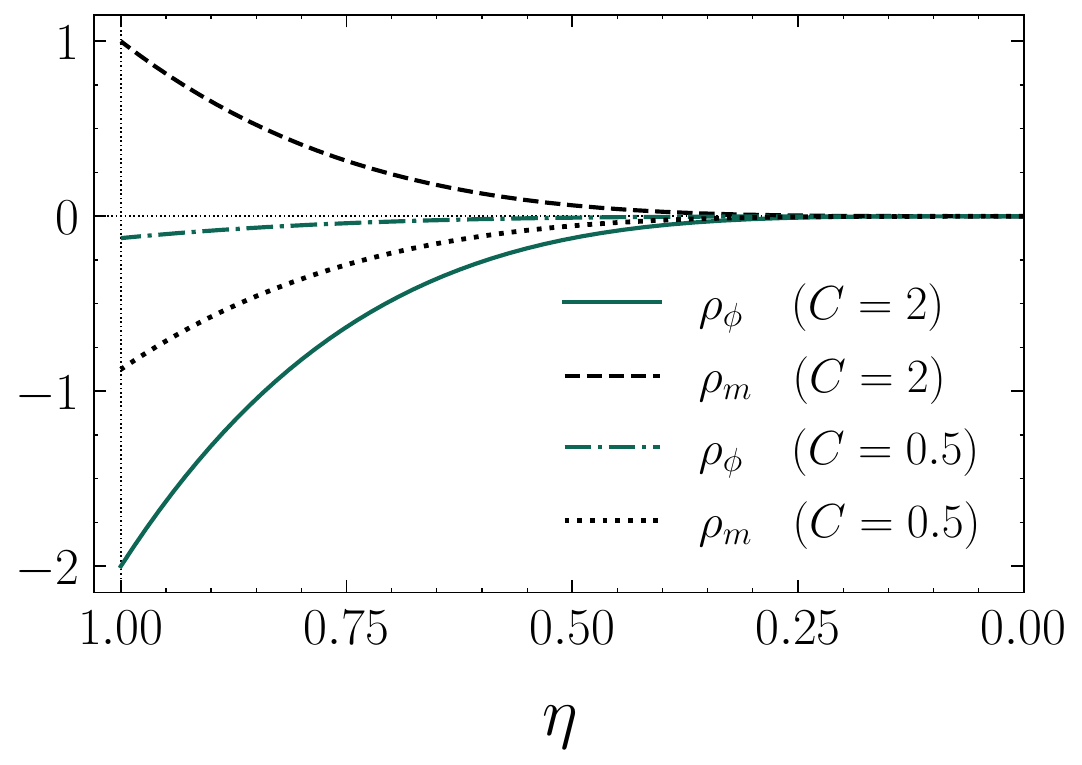}
    \caption{Ghost and matter energy densities for the uncoupled case with different values of $C$.}
    \label{rho_phi_un}
\end{figure}
By introducing a new variable,
\be\label{eta_def}
\eta=\frac{r_0}{r}\,,
\ee
the domain of $r$, {\it i.e.}, ${r\in [r_0,\infty [}$, is compactified to ${\eta\in \,]0,1]}$, which simplifies the analysis of the solutions.

\begin{figure*}[t]
    \centering
    \includegraphics[scale=0.335]{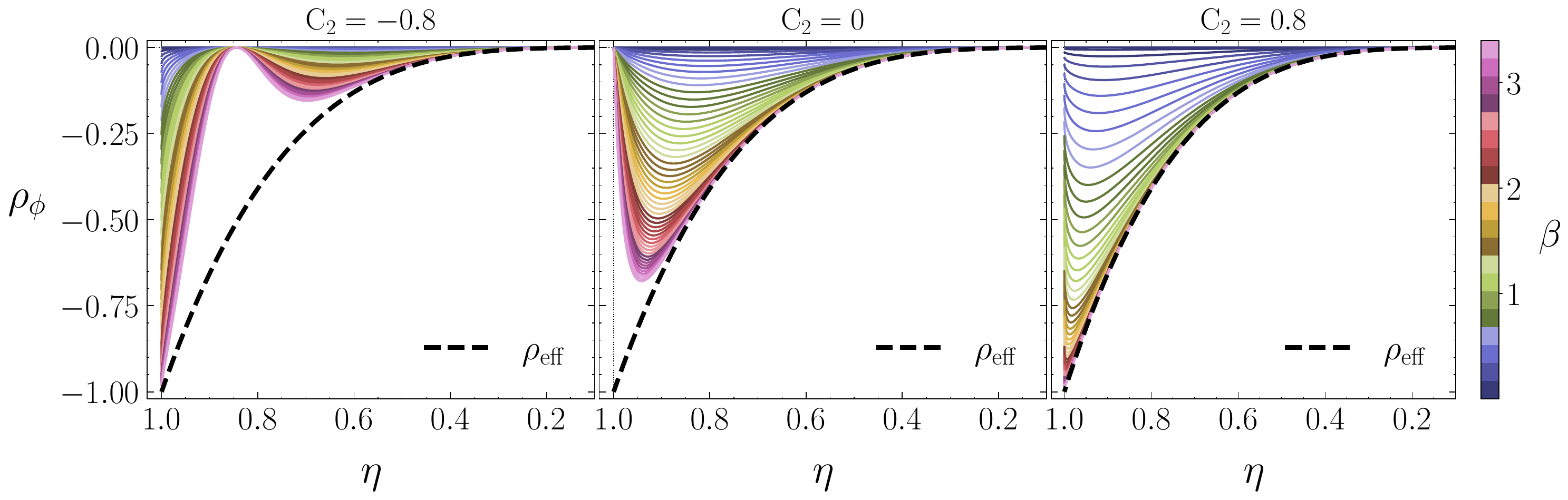}
    \caption{Energy density profile for the ghost scalar field with $V=0$ coupled to matter, given by Eq.~\eqref{analytical_rho_phi} for different values of $\beta$ and ${\rm C}_2$.}
    \label{fig:rho_phi_1}
\end{figure*}
\begin{figure*}[t]
    \centering
    \includegraphics[scale=0.335]{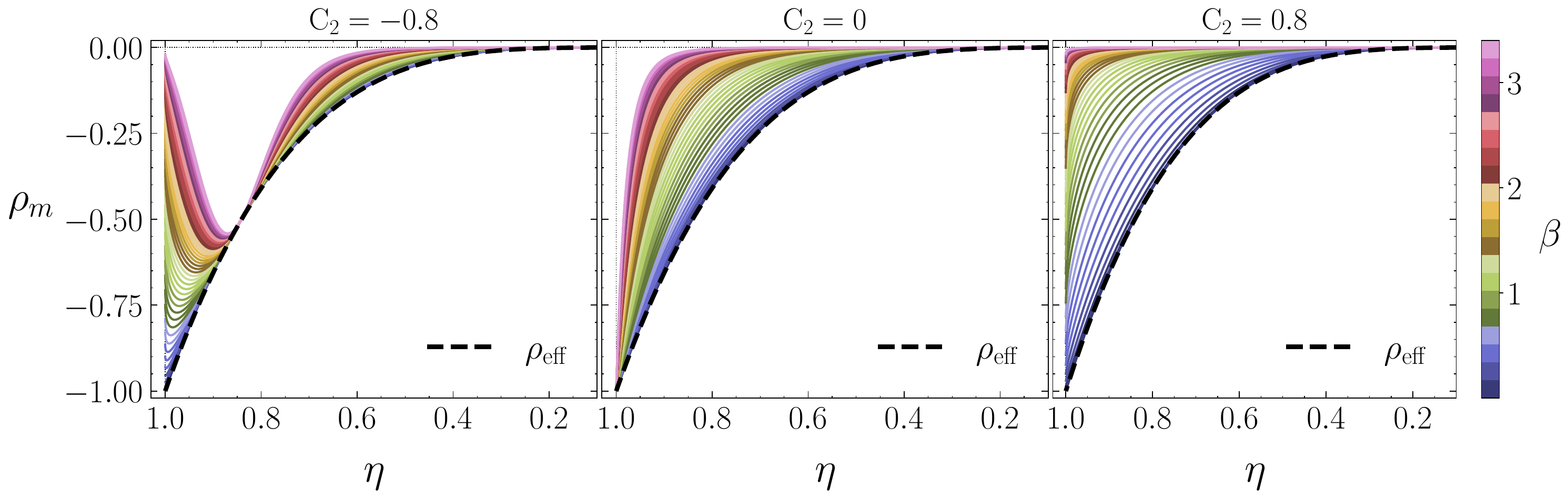}
    \caption{Energy density profile for coupled matter, ${\rho_m=\rho_{\rm eff}-\rho_{\phi}}$ for different values of $\beta$ and ${\rm C}_2$.}
    \label{fig:rho_m_1}
\end{figure*}

With this redefinition at hand, the uncoupled ghost field profile is displayed in Fig.~\ref{phi_un}. Since Eq.~\eqref{uncoupled} is invariant under ${({\phi},C)\mapsto(-{\phi},-C)}$, we restrict ourselves to the ${C\geqslant 0}$ case. Near the throat, the field decays faster as $C$ increases. This can easily be appreciated by expanding Eq.~\eqref{uncoupled} around $r_0$,
\be\phi \approx \phi_0+C\sqrt{2}\sqrt{\frac{r}{r_0}-1}+\mathcal{O}\left[(r-r_0)^{3/2}\right]\,.\ee
The special case $C=0$ yields a constant value for the ghost field, recovering a model where only matter threads the wormhole, {\it i.e.}, ${{\rho}_\phi=0}$ and ${\rho_m \equiv \rho_{\rm eff}}$. On the other hand, as mentioned above, the choice ${C^2=2}$ (solid line in Fig.~\ref{phi_un}) recovers a theory where the ghost field is the sole source supporting the wormhole \cite{Ellis:1973yv,Karakasis:2021tqx}, ${\rho_m=0}$ and ${{\rho}_\phi\equiv\rho_{\rm eff}}$.

The energy densities for both the ghost and the matter species are displayed in Fig.~\ref{rho_phi_un}.
Depending on the value of the integration constant $C$, which by Eq.~\eqref{dif_uncoupled} is directly related with the ghost value at the throat, the energy densities of both sources will contribute to 
sum up to $\rho_{\rm eff}$. However, although the energy density of the ghost, Eq.~\eqref{rho_phi_uncoupled}, is strictly negative, the energy density of matter obviously can attain positive values. Therefore, one may tailor this pure uncoupled solution, through the value of $C$, so as to render the ghost the only exotic source while sustaining the wormhole. Contrary to what concerns matter, the scalar field, if present, is doomed to violate all of the energy conditions due to Eq.~\eqref{nec1}.

\subsection{Haunting matter I: Coupled noninteracting ghost}
\label{SectionIII.B}

The novelty in this scenario is that interestingly we are also able to find an analytical solution to Eq.~\eqref{kleingordon1} for a general $\beta$, in particular,
\begin{eqnarray}
\phi&=&{\rm C}_1- \frac{1}{\beta} \log\Bigg\{\cosh \Bigg[ \sqrt{2}\beta\tan^{-1}\left(\sqrt{\frac{r^2}{r_0^2}-1}\,\right)
    \nonumber \\
&&\hspace{4cm}+\beta\, {\rm C}_2\Big]\Big\},\label{analytical_1}
\end{eqnarray}
which is well behaved in the entire domain of ${r\in[r_0,\infty [}$, with ${\rm C}_1$ and ${\rm C}_2$ being the two integration constants. Note that the constant ${\rm C}_1$ has the same dimensions of the field $\phi$ (mass) and the other integration constant was explicitly written as the product ${\beta\, {\rm C}_2}$; thus, ${\rm C}_2$ is dimensionless. Although Eq.~\eqref{analytical_1} is not strictly defined for ${\beta=0}$, it has a well-behaved uncoupled limit, ${\beta\rightarrow 0}$, in which case the field freezes, ${\phi\rightarrow {\rm C}_1}$, and its energy density $\rho_{\phi}$ would be zero according to Eq.~\eqref{rho_phi}. Consequently, in this limit ${\beta\rightarrow 0}$, only matter threads the wormhole, ${\rho_{\rm eff}\equiv\rho_m}$. Complementarily, in the weak coupling limit, we may expand Eq. \eqref{analytical_1} at first order around ${\beta=0}$, yielding
\be
\phi\approx {\rm C}_1-\beta\left[\tan^{-1}\left(\sqrt{\frac{r^2}{r_0^2}-1}\right)+\frac{ {\rm C}_2}{\sqrt{2}}\right]^2 + \mathcal{O}\left(\beta^2\right)\,,
\ee
which is valid for ${\beta \ll 1}$. In contrast, the strong coupling regime, {\it i.e.}, ${\beta\rightarrow\infty}$, yields ${\rho_{\rm eff}\equiv\rho_{\phi}}$, which reproduces a wormhole threaded only by the scalar source, with vanishing matter, $\rho_m=0$. 

The coupled ghost Eq. \eqref{analytical_1} attains the values
\be\label{dif_coupled0}
\phi_0={\rm C}_1- \frac{1}{\beta} \log\left[ \cosh \left(\beta {\rm C}_2\right)\right]
\ee
at the throat, whereas 
\be\label{dif_coupled}
\phi_{\infty}=\lim_{r\rightarrow\infty}\phi={\rm C}_1-\frac{1}{\beta}\log \left\{\cosh \left[\beta\left(\frac{\pi}{\sqrt{2}} +{\rm C}_2\right) \right]\right\}
\ee
at infinity. Setting $\phi_{\infty}=0$ enables us to find the value for the ${\rm C}_1$ constant given in terms of $C_2$ and $\beta$. Moreover, assuming ${\phi_{\infty}=0}$, the scalar field behavior given by Eq.~\eqref{analytical_1} is symmetric in ${(\phi ,\beta )\mapsto (-\phi , -\beta)}$; thus, henceforth, we will focus solely on the positive coupling case, {\it i.e.}, ${\beta >0}$.
In this scenario, the energy density for the field \eqref{analytical_1} can simply be found through Eq.~\eqref{rho_phi} and reads
\be\label{analytical_rho_phi}
\rho_{\phi}=-\frac{r_0^2}{r^4} \tanh^2 \left[\sqrt{2} \beta \tan^{-1}\left(\sqrt{\frac{r^2}{r_0^2}-1}\,\right)+\beta {\rm C}_2\right]\,,
\ee
which, as desired, vanishes as $r\rightarrow\infty$. At the throat of the wormhole, the ghost energy density above depends on both the values of $\beta$ and ${{\rm C}_2}$, as follows,
\be
\rho_\phi(r_0) = -\frac{\tanh ^2\left( \beta  {\rm C}_2 \right)}{ r_0^2}\,,
\label{rho_phi_throat}
\ee
which vanishes when either one of these values approaches zero. This result is consistent with the fact found in the ${\beta\rightarrow 0}$ limit, namely, that in the weak coupling regime the field vanishes or, more precisely, approaches a constant behavior rendering a vanishing energy density. Along the same lines, the case $\beta\rightarrow\infty$ gives $\rho_{\phi}=\rho_{\rm eff}$ independently of the (nonvanishing) value of ${\rm C}_2$.

In Figs.~\ref{fig:rho_phi_1} and \ref{fig:rho_m_1} we show the ghost and matter energy densities, respectively, for different values of the coupling $\beta$ and ${\rm C}_2$. 

Since the interaction between the species can be understood as an energy transfer between matter and the scalar field, one can already verify from the weak coupling regime, ${\beta\rightarrow 0}$, that all the energy density of the wormhole is contained in matter (${\rho_m\approx\rho_{\rm eff}}$). As $\beta$ grows, energy is transferred from the matter component into the ghost scalar until, in the strong coupling regime, it is practically only the scalar field that threads the wormhole.

 When ${\rm C}_2=0$, the energy density of the ghost scalar vanishes at the throat, in which case it is the matter that sustains the wormhole throat, and thus violates the NEC at $r_0$. When for a fixed $\beta$ we increase ${\rm C}_2$ for positive values, {\it i.e.}, ${{\rm C}_2>0}$, the energy densities approach their limiting behavior more abruptly, thus closer to the throat. 
The case ${\rm C}_2<0$ on the other hand possesses interesting properties. Note that in such cases  
Eq.~\eqref{analytical_rho_phi}
vanishes at some specific radius, $r_{\star}$, expressed in terms of ${\rm C}_2$,
\be\label{r_star}
r_{\star}= r_0 \sqrt{1+\tan^2\left(\frac{{\rm C}_2}{\sqrt{2}}\right)}\,\,.
\ee
This means that once a negative value for ${\rm C}_2$ is chosen,
there is a particular radius, given by Eq.~\eqref{r_star}, such that, at that distance from the throat, the ghost density vanishes, ${\rho_\phi(r_{\star})=0}$. In Figs.~\ref{fig:rho_phi_1} and \ref{fig:rho_m_1}, we show this behavior for an illustrative value of ${\rm C}_2=-0.8$.

From the limiting cases studied above, it is straightforward to show that the energy densities of the ghost and matter are always contained within the regions,
\be
0 > \left\{\rho_\phi\,, \rho_m\right\} > \rho_{\rm eff}\,.
\ee
However, by noticing that for each species $\rho=-\tau$, we gather that both fluids violate all the energy conditions. Due to the latter equality, which is valid provided $V=0$ and $b(r)$ given by Eq.~\eqref{shapefunc}, both the NEC and weak energy condition (WEC) profiles can also be contemplated through Figs.~\ref{fig:rho_phi_1} and \ref{fig:rho_m_1}, as in this specific case the NEC is explicitly given by ${\rho-\tau=2\rho}$.

Note that this framework was attained by fixing ${\phi_\infty =0}$. Nonetheless, given the fact that the field equations
\eqref{ee1}-\eqref{ee3} and \eqref{kleingordon1} depend solely on $\phi'$, 
one may relax this condition since the determination of $\phi_0$, which was automatically determined by
imposing the vanishing of $\phi_{\infty}$, does not affect these equations.
This can be understood by noticing that from Eq.~\eqref{dif_coupled} onward we have fixed ${\rm C}_1$ so as ${\phi_\infty=0}$. 
However, the ghost energy density \eqref{analytical_rho_phi} does not depend on ${\rm C}_1$.
If the $\phi_\infty$ value is not set to zero,
the ghost field $\phi$ would simply freeze to a constant limiting value
far from the wormhole, with no influence on the dynamics whatsoever.

\section{Solutions with a scalar potential}\label{sec_mass}

Let us now turn our attention to the case where a scalar self-interacting potential is present in the total action \eqref{action}, {\it i.e.}, ${V\neq 0}$. In such a case, our approach will be different from the one employed in Sec.~\ref{sec_massless}. Instead of solving the Klein-Gordon equation with respect to the field, we shall assume specific dynamical behaviors for both the ghost and for the metric functions and find the particular self-interacting potential $V$ that solves the gravitational scheme. In particular, let us assume that the scalar field follows a simple decaying form throughout the wormhole,
\be\label{field_ansatz}
\phi = \phi_0\left(\frac{r_0}{r}\right)^{\alpha}\,,
\ee
with ${\alpha >0}$ being a constant. For the metric functions, let us write the following
ansätze,
\be\label{metric_funcs}
b=r_0\left(\frac{r_0}{r}\right)^{\gamma}\,,\quad\text{and}\quad \Phi=\Phi_0\left(\frac{r_0}{r}\right)^{\lambda}\,,
\ee
with constant ${\gamma >-1}$ such that $b(r)$ obeys the flaring-out condition, and a non-negative constant value of $\lambda$ so the redshift function $\Phi$ vanishes as ${r\rightarrow \infty}$. 

\subsection{Uncoupled self-interacting case}\label{uncoupled_pot}

Under the hypotheses above,
the Klein-Gordon equation for the ghost, Eq.~\eqref{kleingordon}, in the noninteracting case, {\it i.e.}, ${\beta =0}$, becomes
\begin{eqnarray}
\frac{\dd V}{\dd r}&=&\frac{\alpha ^2 \phi_0^2}{r^3} \left(\frac{r_0}{r}\right)^{2\eta } \left\{ \eta-1 +\lambda  \Phi_0 \left(\frac{r_0}{r}\right)^{\lambda }\,\right.\nonumber \\
&\,\,&  -\left. \left(\frac{r_0}{r}\right)^{\gamma +1} \left[\frac{\gamma}{2} + \eta -\frac{1}{2} + \lambda  \Phi_0 \left(\frac{r_0}{r}\right)^{\lambda }\right]\right\}\,,\hspace{0.8cm}\label{kg_2_un} 
\end{eqnarray}
which yields the following analytical solution,
\begin{figure}[t]
    \centering
    \includegraphics[scale=0.415]{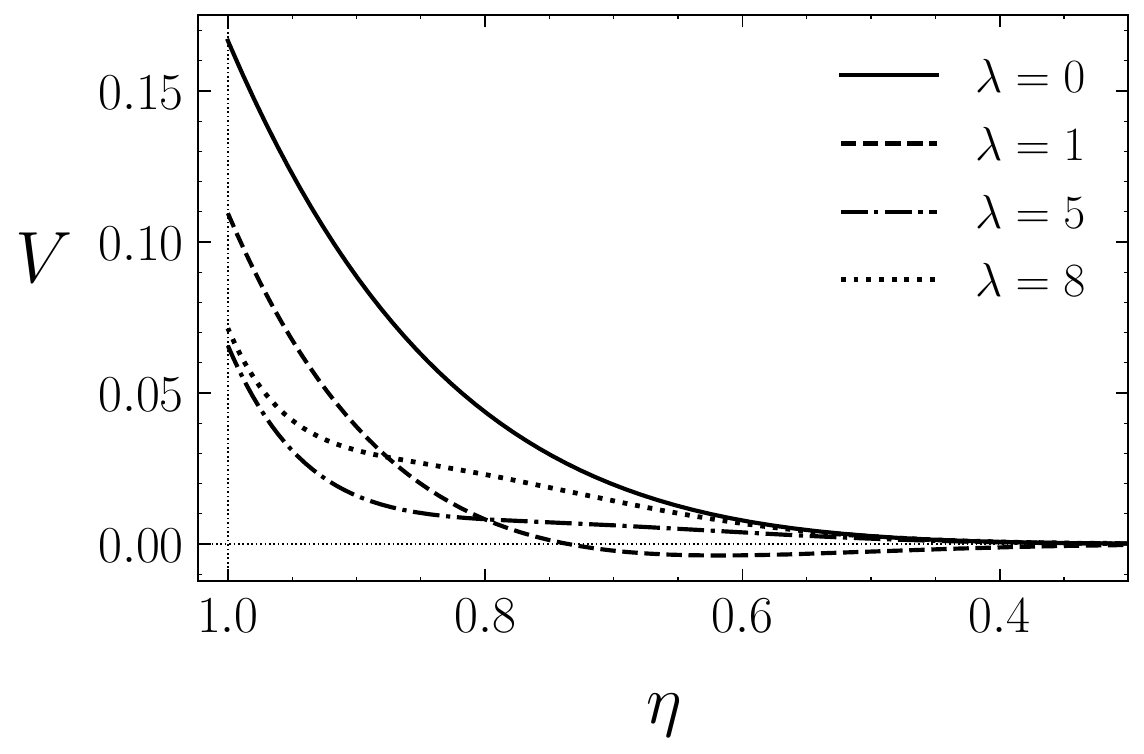}
    \caption{Scalar field potential given by Eq.~\eqref{sol_pot} with $\alpha=\gamma=\phi_0=\Phi_0=1$ and different values of $\lambda$.}
    \label{fig:pot}
\end{figure}
\begin{eqnarray}\label{sol_pot}
V &=& \frac{\alpha^2\phi_0^2}{2r^2}\left(\frac{r_0}{r}\right)^{2\alpha}\left\{ \frac{1-\alpha}{1+\alpha}+\left(\frac{r_0}{r}\right)^{\gamma +1}\frac{2\alpha +\gamma -1}{2\alpha + \gamma +3}\right.\nonumber \\
&& +\, 2\lambda\Phi_0\left[ \left(\frac{r_0}{r}\right)^{\gamma +\lambda +1}\frac{1}{2\alpha+\gamma+\lambda+3}\right. \nonumber \\
&& -\left. \left.\left(\frac{r_0}{r}\right)^{\lambda}\frac{1}{2\alpha+\lambda+2}\right]
 \right\}+V_{\infty}\,, \label{pot}
\end{eqnarray}
\begin{figure}[t]
    \centering
    \includegraphics[scale=0.45]{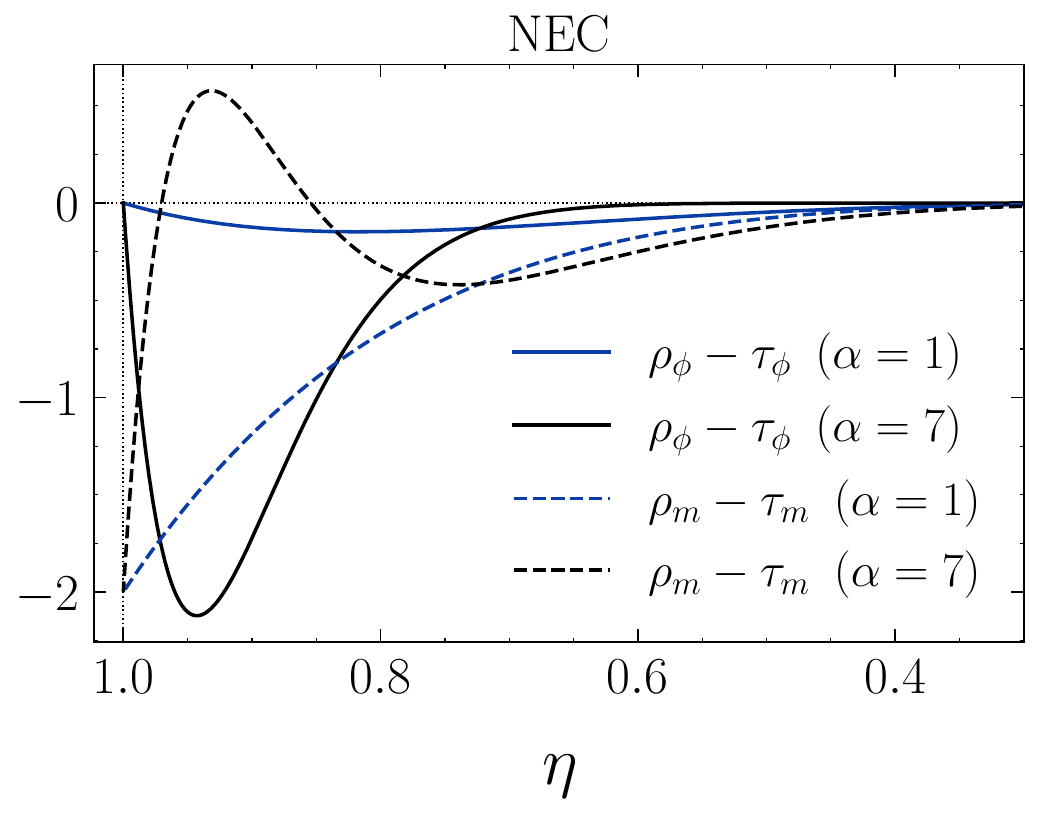}
    \caption{Null energy condition profile for the scalar field (solid) and matter (dashed), for the solution presented in Sec.~\ref{uncoupled_pot}, with a constant redshift function, $\lambda = 0$, $\gamma=\phi_0=1$ and different values of $\alpha$.}
    \label{fig:nec_uncoupled_pot}
\end{figure}
with $V_{\infty}$ being an integration constant standing for the value of the potential at infinity, which we set to zero in order for the energy density of the ghost to vanish at ${r\rightarrow\infty}$. It is straightforward to get an expression for $V(\phi)$ by expressing Eq.~\eqref{field_ansatz} in the form $r(\phi)$
and plugging the result into Eq.~\eqref{pot}. Therefore, due to our specific choices for the field, Eq.~\eqref{field_ansatz}, and metric functions, Eq.~\eqref{metric_funcs}, we notice that $V$ follows a simple sum of powers of $\phi$. The simple case with ${\Phi_0=0}$ results in the following potential
\be V(\phi)\propto A\,\phi^{\,2\left(1+\frac{1}{\alpha}\right)}+B\,\phi^{\,2+\frac{3+\gamma}{\alpha}}\,,
\ee
where $A$ and $B$ are constants which depend on the free parameters, namely, $r_0$, $\Phi_0$, $\phi_0$, $\alpha$, and $\gamma$. The $\Phi_0\neq 0$ case is depicted in Fig.~\ref{fig:pot} for different values of $\lambda$. Note that the potential of the field will always contribute to the total energy of the ghost at the throat, except in the specific case where
\be
\Phi_0 = \frac{\left( 2\alpha + \lambda + 2 \right)\left( 2\alpha+\gamma+\lambda \right)}{\lambda\left( 1+\alpha \right)\left( 2\alpha + \gamma +3 \right)}\,,
\ee
in which case the total potential vanishes at $r_0$.

In all these cases of this section, Sec.~\ref{uncoupled_pot}, the NEC for the scalar field does not depend on the redshift function parameters, which is transparent from Eqs. (\ref{rho_phi}) and (\ref{tau_phi}) that provide $\rho_\phi-\tau_\phi = - (1-b/r)\phi'^{2}$, and thus reads
\be\label{nec_field_pot_uncoupled}
\rho_{\phi}-\tau_{\phi} = -\frac{\phi_0\alpha^2}{r^2}\left(\frac{r_0}{r}\right)^{2\alpha}\left[1-\left(\frac{r_0}{r}\right)^{1+\gamma}\right]\,,
\ee
which is always nonpositive and vanishes at the throat. 
Indeed, as $\rho_\phi-\tau_\phi \rightarrow 0$, at the wormhole throat, we have $\rho_{m} - \tau_{m} < 0$ at $r=r_0$. Thus, the NEC is generically violated at the wormhole throat by the matter threading the wormhole for the case analyzed in this section.
This means that, although the ghost scalar field will violate the classical energy conditions throughout the wormhole spacetime, at the throat, $r=r_0$, the ordinary matter source will be responsible for generating a tension (negative radial pressure) larger than the energy density, {\it i.e.}, ${\tau_m>\rho_m}$, in order to hold the wormhole throat open. On the other hand, it is possible to tune the parameters such as to minimize the NEC violation close to the throat. This can be observed in Fig.~\ref{fig:nec_uncoupled_pot}, in which by increasing the decaying power of the field, that is, $\alpha$, it is possible for ordinary matter to obey the NEC for some interval of $r$ during which it is the ghost field that violates the NEC. We verified that this behavior can be more prominent when allowing other parameters, in particular $\gamma$ and $\phi_0$, to vary. 
However, as mentioned, at the throat, we always have ${\rho_\phi-\tau_\phi=0}$.

\subsection{Haunting matter II: Coupled self-interacting ghost}\label{coupled_pot}

We now consider the general case where the ghost field has a self-interacting potential, $V\neq 0$, and couples to matter, $\beta\neq 0$. In order to illustrate the rich phenomenology of this scenario and due to the complicated form of the differential equation to solve, we will focus on the specific case where the scalar field follows an inverse radial function ${\phi=\phi_0 r_0/r}$, {\it i.e.}, Eq.~\eqref{field_ansatz} with $\alpha=1$ and a well-known wormhole shape function $b=r_0^2/r$. Regarding the redshift function, we will assume for simplicity that it depends explicitly on the coupling constant $\beta$ as
\be\label{redshift_ansatz}
\Phi = \Phi_0\left(\frac{r_0}{r}\right)^{\beta}\,.
\ee
Thus, the redshift of photons and tidal accelerations felt by a traveler will be directly influenced by the interaction between the ghost and matter. 

Within this setting, we are able to find an analytical expression for the potential. However, due to its long expression we chose to write it simply as\footnote{Here, $\Gamma (a,z)$ denotes the incomplete gamma function,
$\Gamma (a,z)=\int_z^{\infty}x^{a-1}{\rm e}^{-x}\dd x\,.$
}
\be\label{sol_pot_2}
V(\phi)=\sum_{i=0}^6\Bigl\{{A_{(i)}}\phi^i+{\rm e}^{B\phi}\bigl[C+D_{(i)}\Gamma\left(E_{(i)},B\phi\right)\,\bigr]\Bigr\}\,,
\ee 
where $A_{(i)}, B, C, D_{(i)}$, and $E_{(i)}$ are constants depending on $\phi_0, \Phi_0$, and $\beta$ and with the real domain on ${\beta\phi_0<0}$. The integration constant was fixed such that ${V\rightarrow 0}$ as ${r\rightarrow\infty}$.

\begin{figure}[t]
    \centering
    \includegraphics[scale=0.4]{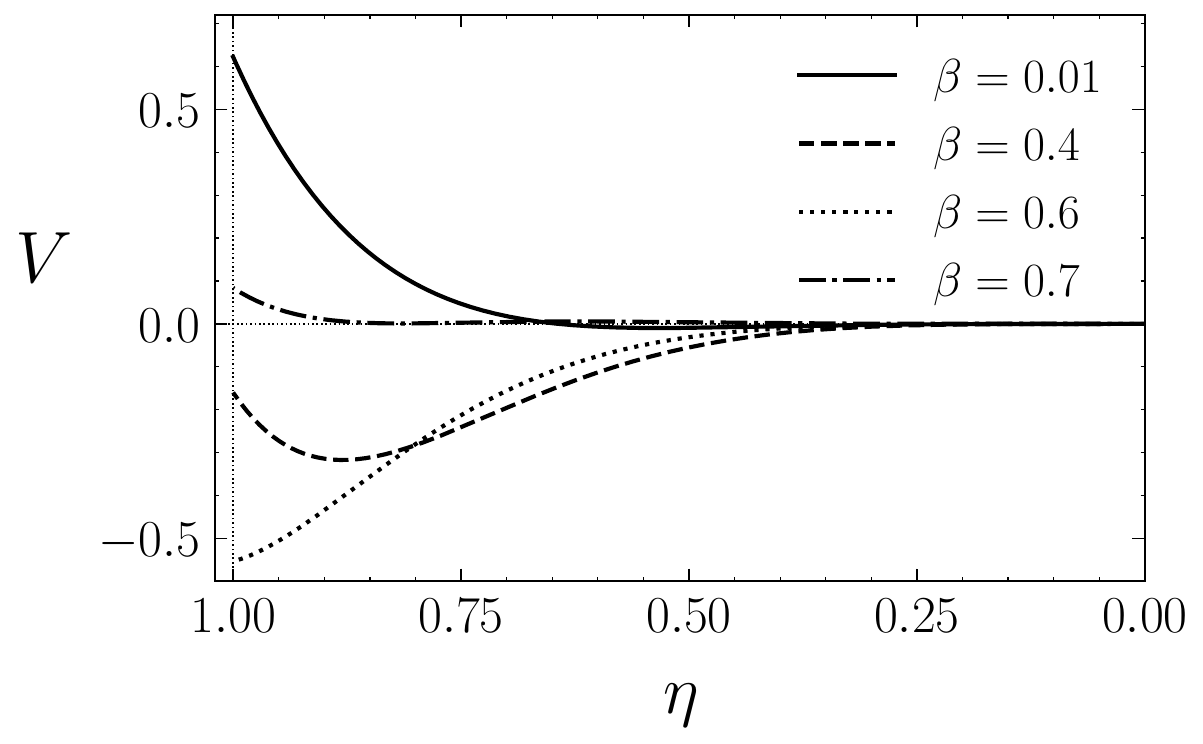}
    \caption{Scalar field potential given by Eq.~\eqref{sol_pot_2} with $\Phi_0=3$, $\phi_0=-2$, and different values of the coupling $\beta$.}
    \label{fig:pot2}
\end{figure}

\begin{figure}[t]
    \centering
    \includegraphics[scale=0.45]{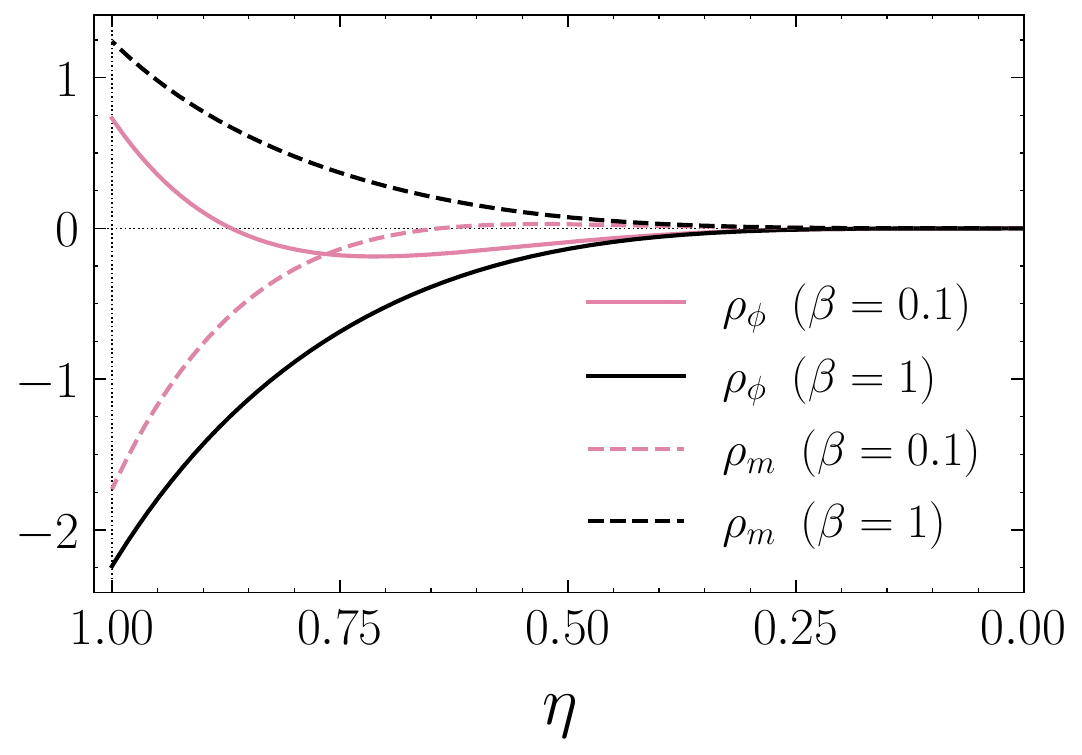}
    \caption{Energy density profile for the scalar field (solid) and matter (dashed), for the solution presented in Sec.~\ref{coupled_pot}, with ${\Phi_0=1}$, $\phi_0=-2$, and different values of the coupling $\beta$.}
    \label{fig:rhos_coupled_pot}
\end{figure}

For illustrative purposes, we plot the scalar self-interacting potential in Fig.~\ref{fig:pot2}. Due to the specific forms of the energy density and tension of the field, given by Eqs.~\eqref{rho_phi} and \eqref{tau_phi}, respectively, the difference $\rho_\phi-\tau_\phi$  does not depend on the specific shape of the potential nor, thus, the NEC. Therefore, the combination $\rho_\phi-\tau_\phi$ follows the same profile as in the previous subsection given by Eq.~\eqref{nec_field_pot_uncoupled} with ${\alpha=\gamma=1}$. The same conclusions are thus drawn; at (and near) the throat, it is matter that violates the energy conditions, thus sustaining the wormhole. Nonetheless, our solutions allow for a positive energy density of matter in the whole interval of values of $\eta$ depending on the parameters. This trend is depicted in Fig.~\ref{fig:rhos_coupled_pot} in which for a stronger interaction, {\it i.e.}, higher values of $\beta$, the value of the energy density ${\rho_\phi(r_0)}$ at the throat decreases, resulting in a positive matter energy density. We emphasize the fact that this trend is dependent on the choices for $\phi_0$ and $\Phi_0$.

The aforementioned traversability conditions in Sec.~\ref{wormhole:metric} (the finite character of $\Phi$) render the wormhole free from horizons and singularities, thus having a traversable nature. However, for a trip to be within human reach, some extra conditions should be imposed on the metric functions \cite{Morris:1988cz}. As mentioned before, since we have tailored the redshift function to explicitly depend on the parameter $\beta$, this will affect the forces felt by a traveler when traversing the wormhole. Although an advanced civilization might have the technology to withstand large G forces, that is (still) not the case for human beings. Thus, for this present wormhole, it is of paramount importance that the acceleration felt by a human traveler, when radially traversing the wormhole (see Ref. \cite{Alcubierre:2017pqm} for more details), given by
\be
|a|=\left| \left(1-\frac{b(r)}{r}\right)^{1/2} {\rm e}^{-\Phi(r)}\left(\gamma_L {\rm e}^{\Phi(r)}\right)' \right|\leqslant g_{\oplus}\,,
\ee
\begin{figure}[t]
    \centering    \includegraphics[scale=0.4]{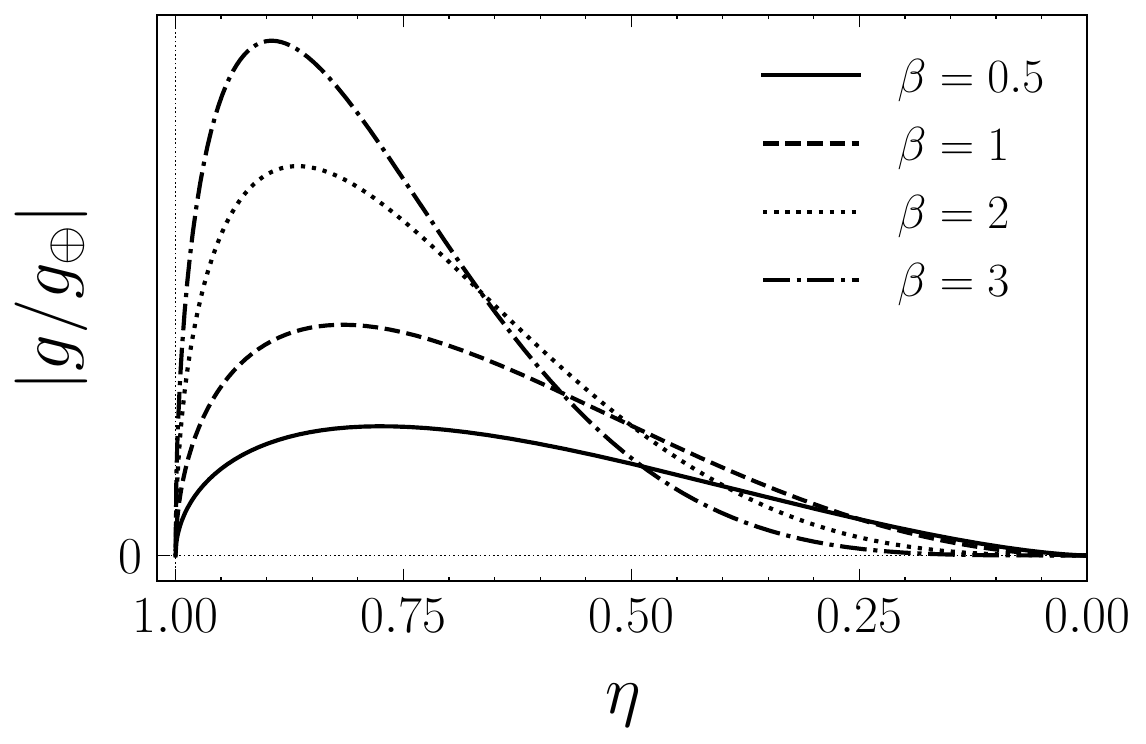}
    \caption{Impact of the coupling on the acceleration felt by a traveler, Eq.~(\ref{accel}), for a fixed value of $\Phi_0$ and different values of the coupling $\beta$.}
\label{fig:acceleration}
\end{figure}
is less than Earth's gravitational field. In the above equation, ${\gamma_L=\left(1-v^2/c^2\right)^{-1/2}}$ is the Lorentz factor, $v$ being the velocity of the traveler, and $g_{\oplus}$ denotes Earth's gravitational force. Assuming a human is traveling with a nonrelativistic speed, $\gamma\approx 1$, for this present solution, with
${\phi=\phi_0 r_0/r}$, ${b=r_0/r^2}$ together with
Eq.~\eqref{redshift_ansatz}, we have
\be
\frac{g}{g_{\oplus}}=-\beta\frac{\Phi_0}{r_0}\eta^{1+\beta}\sqrt{1-\eta^2}\,.
\label{accel}
\ee
Note that negative or positive values express the attractive or repulsive character of the wormhole. The above equation tells us that, given a wormhole with a large $\Phi_0$, the interaction, provided by $\beta$, between the fluid species has to be small for human traversability. On the other hand, a strong interaction needs to be compensated with small values of $\Phi_0$. This behavior merely expresses the smallness conditions for Eq.~\eqref{redshift_ansatz}. The influence of the interaction on the acceleration felt by a traveler is depicted in Fig.~\ref{fig:acceleration}. While for small values of $\beta$ the increase goes as a multiplicative factor $\beta$, as the coupling becomes stronger, the term $\eta^{1+\beta}$ dominates, which is evident away from the throat since $a(r_0)=0$, generating a faster decrease of the profile, resulting on negligible Gs away from the throat, or at least smaller than for very small couplings.

\section{Conclusions and outlook}\label{sec:conclusions}

In this work we have obtained exact solutions for wormhole geometries supported by a ghost scalar field that is conformally coupled to an anisotropic distribution of matter. The main features of the theory were explored, such as the scalar field phenomenology, the matter threading the wormhole, and the role played by the nonminimal coupling. We verified that for a vanishing self-interaction potential, with a matter-scalar noninteraction, although the energy density of the ghost is strictly negative, the energy density of matter attains positive values. This effect was also shown to be possible for the coupled self-interacting case. Furthermore, specific traversability conditions were analyzed for coupled massive ghosts, in the presence of a scalar potential. This extends several works already presented in the literature by allowing for a coupling between the scalar source and matter, in the Einsten frame. 

The geometrical aspects of conformal mappings in wormhole and black hole physics were explored in Refs.~\cite{Faraoni:2015paa,Hammad:2018ldj}. However, studies of wormholes within STTs in the literature \cite{Bronnikov:2001ae,Bronnikov:2010tt,Darabi:2010je,Faraoni:2016ozb,Barcelo:1999hq,Hohmann:2018shl} neglect the extra matter source which couples to the scalar degree of freedom under the Weyl scaling \eqref{weyl}, and most conduct the analysis in the Jordan frame. Here, we have obtained exact solutions considering a general matter source; contemplated the scalar self-interacting case; and, additionally, by doing the analysis in the Einstein frame, were able to explicitly trace the effects arising from the interaction, which becomes evident in this gravitational frame.

An interesting extension of these static and spherically symmetric wormholes would be to include dynamic geometries \cite{Kar:1994tz} and rotating solutions \cite{Teo:1998dp}. These have been known to improve the violation of the energy conditions; in particular, for rotating wormhole solutions, the exotic matter is distributed in specific regions, such that a radially infalling observer may avoid the NEC violating matter altogether, and the dynamic geometries may satisfy the energy conditions for specific finite time intervals \cite{Kar:1994tz}. In this realm, stability issues are also of fundamental importance, and it was shown that wormholes in specific modified theories of gravity are linearly stable with respect to radial perturbations \cite{Kanti:2011yv}. Thus, extensions of the solutions outlined in this work to the rotating and dynamic solutions would provide further insight into these extremely interesting geometries. Work along these lines is presently underway.


\acknowledgments

B.J.B. acknowledges support from the Funda\c{c}\~{a}o para a Ci\^{e}ncia e a Tecnologia (FCT) through the project BEYLA: BEYond LAmbda with reference number PTDC/FIS-AST/0054/2021 and the South African NRF Grants No. 120390 (reference: BSFP190416431035) and No. 120396 (reference: CSRP190405427545).
A.d.l.C.D. acknowledges further support from University of Cape Town
URC Grant 2022 and Grant No. PID2021-122938NB-I00 funded by MCIN/AEI/
10.13039/501100011033 and by “ERDF A way of making Europe” and BG20/00236 action (Ministerio de Universidades, Spain). 
F.S.N.L. acknowledges support from the Funda\c{c}\~{a}o para a Ci\^{e}ncia e a Tecnologia (FCT) Scientific Employment Stimulus contract with reference CEECINST/00032/2018, from the FCT research grants UIDB/04434/2020 and UIDP/04434/2020, and through the FCT Projects No. PTDC/FIS-AST/0054/2021 and No. CERN/FIS-PAR/0037/2019. 

\bibliography{bib}

\end{document}